\documentclass[prl,twocolumn,showpacs]{revtex4}
\usepackage{bm}
\usepackage{graphicx}

\begin{document}
\title{The spherical $2+p$ spin glass model: an exactly solvable
model for glass to spin-glass transition.}

\author{A. Crisanti}
\email{andrea.crisanti@phys.uniroma1.it}

\author{L. Leuzzi}
\email{luca.leuzzi@roma1.infn.it}

\affiliation{Dipartimento di Fisica, Universit\`a di Roma ``La Sapienza''}
\affiliation{Istituto Nazionale Fisica della Materia, Unit\`a di Roma, 
             and SMC,
             P.le Aldo Moro 2, I-00185 Roma, Italy}

\begin{abstract}
We present the full phase diagram of the spherical $2+p$ spin glass model
with $p\geq 4$. The main outcome is the presence of a new phase 
with both properties of Full Replica Symmetry Breaking (FRSB) phases 
of discrete models, e.g, the Sherrington-Kirkpatrick model, and
those of One Replica Symmetry Breaking (1RSB). The phase, which separates 
a 1RSB phase from FRSB phase, is
described by an order parameter function $q(x)$ with a continuous part 
(FRSB) for $x<m$ and a discontinuous jump (1RSB) at $x=m$.
This phase has a finite complexity which leads to different dynamic and static 
properties.
\end{abstract} 


\maketitle

%{\bf Introduction}\\
In the last years many efforts have been devoted to the understanding of
complex systems such as spin glasses, structural glasses
and others. Common denominator of all these systems is a large number 
of stable and metastable states \cite{BOOK} whose complex structure 
determines the their static or dynamic behaviors. In this framework mean-field
models, and among them spherical models, represent a valuable tool of 
analytical and theoretical investigation  since they can be largely solved. 
Up to now only spherical models with One Replica Symmetry Breaking (1RSB)
phases were studied, mainly due to 
their relevance for the fragile glass transition \cite{KirThi87,CriSom92,REV}. 

To our knowledge the possibility infinite or Full Replica Symmetry 
Breaking (FRSB) phases in spherical models was
first pointed out by Nieuwenhuizen \cite{Nieuwenhuizen95} on the basis of the
similarity between the replica free energy of some spherical models with 
multi-spin interactions and the 
relevant part of the free energy of the Sherrington-Kirkpatrick (SK) model
\cite{BraMor78,PytRud79}. In this paper Nieuwenhuizen presented some results for
the FRSB phase but a complete analysis was not provided. The problem was 
considered some years later \cite{CiuCri00} in connection with 
the possible different scenarios for the critical dynamics near the glass 
transition \cite{GoeSjo89}. This paper, however, analyzed only 
the dynamical behavior in the 1RSB phase.

{\it The Model}.
The model we shall consider is 
the spherical $2+p$ spin glass model without external field 
defined by the Hamiltonian
\begin{equation}
{\cal H} =  \sum_{i<j}J^{(2)}_{ij}\sigma_i\sigma_j
           +\sum_{i_1<\ldots <i_p}J^{(p)}_{i_1\ldots i_p}
	   \sigma_{i_1}\cdots\sigma_{i_p}
\end{equation}
where $ J^{(p)}_{i_1 i_2..i_p}$ are uncorrelated zero mean 
Gaussian variables of variance
\begin{equation}
\label{varVp}
   \overline{\left(J^{(p)}_{i_1 i_2..i_p}\right)^2} = 
    \frac{J_p^2 p!}{2N^{p-1}}, \qquad i_1 < \cdots < i_p
\end{equation}
and $\sigma_i$ are $N$ continuous variables obeying
the spherical constraint $\sum_i \sigma_i^2 = N$. The properties of the model
strongly depend on the value of $p$: for $p=3$ the model 
reduces to the usual spherical $p$-spin spin glass model in a 
field \cite{CriSom92} with a low temperature 1RSB phase,
while for $p\geq 4$ the model posses an additional
FRSB low-temperature phase \cite{Nieuwenhuizen95}. 
A partial analysis of the phase space of the model 
was carried out in Ref. \onlinecite{CiuCri00} where, however, only the 
dynamical stability of the 1RSB phase was considered leaving out a large 
part of the phase space and in particular the question of the transition
between the 1RSB and the FRSB phases. 

In this Letter we complete the study
of the phase space focusing in particular on the transitions lines.
We have studied the model by three complementary approaches. The first 
employs the replica method and analyze the disorder-averaged logarithm of
the partition function following Ref. \onlinecite{CriSom92}. The second 
approach stars form the microscopic dynamics and extend the results of
Ref. \onlinecite{CiuCri00} while  the latter uses the Thouless-Anderson-Palmer
approach \cite{ThoAndPal77}. In this Letter we shall mainly follow
the replica approach, discussing differences with other approaches when 
necessary. A complete analysis of the properties of the model is
beyond the scope of the Letter and will be presented elsewhere.  

Applying the standard replica method the free energy per spin $f$ can be 
written as function of the symmetric $n\times n$ replica overlap matrix 
$Q_{\alpha\beta}$ as \cite{CriSom92}
\begin{equation}
\label{eq:free1}
-\beta f = -\beta f_0 + s(\infty) + \lim_{n\to\ 0} \frac{1}{n} \max_{Q} G[Q]
\end{equation}
where $f_0$ is an irrelevant  constant, $s(\infty) = (1+\ln 2\pi) / 2$ the 
entropy per spin at infinite temperature $T = 1 / \beta$,
\begin{equation}
\label{eq:free2}
G[Q] = \frac{1}{2} \sum_{\alpha\beta}^{1,n} g(Q_{\alpha\beta}) + 
       \frac{1}{2}\ln\det Q
\end{equation}
and
\begin{equation}
\label{eq:free3}
g(x) = \frac{\mu_2}{2} x^2 + \frac{\mu_p}{p} x^p.
\end{equation}
where we have used the shorthand $\mu_p = (\beta J^{(p)})^2 / 2 p$. 
The spherical constraint is ensured by the condition 
$Q_{\alpha\alpha} = \overline{q} = 1$.

Following Parisi \cite{Parisi80} the overlap matrix $Q_{\alpha\beta}$ 
for a number $R$ of steps in the replica symmetry breaking 
is divided into successive boxes of decreasing size $p_r$, 
with $p_0 = n$ and $p_{R+1}=1$. The replica symmetric case and the 
FRSB case are obtained for $R=0$ and $R\to\infty$, respectively.
In the Parisi scheme the elements of $Q_{\alpha\beta}$ are
then given by
\begin{equation}
Q_{\alpha\beta} \equiv Q_{\alpha\cap\beta=r} = q_r, \qquad r = 0,\cdots, R+1
\end{equation}
with $Q_{R+1} = \overline{q}$, 
where the notation 
$\alpha\cap\beta=r$ means that $\alpha$ and $\beta$ belong to the 
same box of size $p_r$ but to two {\it distinct} boxes of size $p_{r+1} < p_r$.
The matrix obtained is conveniently expressed using the function
\begin{equation}
x(q) = p_0 + \sum_{r=0}^{R} (p_{r+1} - p_r)\, \theta(q - q_r)
\end{equation}
which equals the fraction of pair of replicas with overlap less or 
equal to  $q$. Inserting this structure into eqs. 
(\ref{eq:free1})-(\ref{eq:free3}), neglecting terms of order $O(n^2)$, and
replacing the sums by integrals, one gets after a little of algebra,
\begin{eqnarray}
-2\beta f &=& 2 s(\infty) - 2\beta f_0 
            + \int_{0}^{1} dq x(q)\, \frac{d}{dq}g(q) 
\nonumber\\
	    && + \ln\left(1 - q(1)\right)
	    + \int_{0}^{q(1)} \frac{dq}{\int_{q}^{1} dq'\, x(q')}
\label{eq:rpl-f}
\end{eqnarray}
where $q(1) = q_R$ and $q(x)$ is the inverse of $x(q)$. 
Maximization of $f$ with respect to $q(x)$
leads to the self-consistent equation(s) for the order parameter function 
$q(x)$.
Depending on the value of the coupling strengths $J^{(p)}$ and of the 
temperature $T$ the function $q(x)$ displays different forms
which characterize the different phases of the model.
\begin{figure}
\includegraphics[scale=1.0]{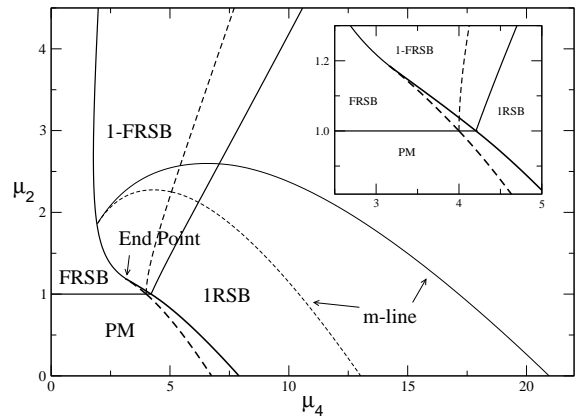}
\caption{Phase digram of the spherical $2+4$ spin glass model.
PM: Paramagnetic phase; 1RSB: one-replica symmetry breaking phase;
FRSB: full replica symmetry breaking phase; 1-FRSB: one-full
replica symmetry breaking. The dashed lines refers to dynamics. 
The $m$-line shown are obtained for $m=0.5$. The continuous transition
between the PM and the FRSB phases and between the FRSB and 1-FRSB phases are
the same for statics and dynamics. Inset: the discontinuous 
transition between the FRSB and the 1-FRSB phases.
}
\label{fig:phdi.2+4}
\end{figure}
Figure \ref{fig:phdi.2+4} shows the phase diagram in the space of the
``natural'' parameters $\mu_p-\mu_2$ for $p=4$. In the following we shall 
limit ourself to the case $p=4$, however the results are qualitatively 
valid for any  $p\geq 4$. The analysis of the figure reveals 
four different phases, which will be discussed in the forthcoming part of this 
Letter.

{\it The Paramagnetic phase} (PM).
This phase exists for not to large values of
coupling parameters strengths and/or high temperature and 
is characterized by a null order parameter function:
$q(x) = 0$ in the whole range $x\in[0,1]$.
The phases becomes unstable
above the line $\mu_2=1$ (DeAlmeida-Thouless line) where the
``replicon'' $\Lambda = 1-\mu_2$ becomes negative. In this region
for $p\geq 4$ and $\mu_p$ not too large a 1RSB solution is also 
unstable and a more complex phase (FRSB) appears.
Below $\mu_2=1$ the PM phase remains stable for all values of $\mu_p$,
similarly to what happens in the spherical $p$-spin model without a field
\cite{CriSom92}, however as $\mu_p$ increases
a more thermodynamically favorable 1RSB phase with a non 
vanishing order parameter appears. 

{\it The One Replica Symmetry Breaking phase} (1RSB).
This phase is characterized by a step-like order parameter function
$q(x) = q_1 \theta(x-m)$ \cite{note1} and is stable as long as 
the replicon eigenvalue is positive:
\begin{equation}
\label{eq:1rsb-stab}
\frac{1}{(1-q_1+m q_1)^2} - \left.\frac{d^2}{dq^2}g(q)\right|_{q=0} > 0
\end{equation}
Maximization of $f$ with respect to 
$q_1$ and $m$ leads to the 1RSB equations whose solution 
can be conveniently expressed 
defining $q_1 = (1-y) / (1- y + my)$ 
in term of the function
\begin{equation}
\label{eq:CS-z}
z(y) = -2y \frac{1-y + \ln y}{1-y}
\end{equation}
introduced in Ref. \onlinecite{CriSom92} for the solution of the 
spherical $p$-spin spin glass model. For $p=4$ the solution reads
\begin{eqnarray}
\label{eq:1rsb}
\mu_4 &=& 2 (1-z(y))  \frac{(1-y+my)^4}{m^2 y (1-y)^2} \\
\mu_2 &=& (2z(y) - 1) \frac{(1-y+my)^2}{m^2 y } \nonumber
\end{eqnarray}
By fixing the value of $m\in[0,1]$ and
varying $y$ from $y_{\min}: z(y_{\min}) = 1/2$ ($\mu_2=0$) to 
$y_{\max}: z(y_{\max}) = 1/2(1+y_{\max})$, where the replicon 
(\ref{eq:1rsb-stab}) vanishes, one obtains the so called $m$-lines.
The transition between the PM and the 1RSB phases corresponds to the 
$m=1$-line. Along this line $q_1$ jumps discontinuously from zero (PM) to
a finite value (1RSB) however, since $m=1$, the thermodynamic quantities remain
continuous. Inserting into (\ref{eq:1rsb}) the value $y_{\max}$, for which the
replicon vanishes, and varying $m$ from $1$ to $0$ one obtains the critical 
line between the 1RSB and the 1-FRSB phase.

The static approach requires that $f$ be maximal with 
respect to variations of $m$. The dynamics, on the other hand, leads to
the different conditions (marginal condition) 
\begin{equation}
\label{eq:1rsb-marg}
\frac{1}{(1-q_1)^2} - \frac{d^2}{dq_1^2}g(q1) = 0
\end{equation}
which can be stated by saying that the derivative of $f$ with respect to 
$m$ (the complexity) be maximal 
\cite{CriHorSom93,KurParVir93,CriSom95,Monasson95}.
As a consequence the transition lines for dynamics and statics do not coincide.
Due to space limitations the equations for the dynamical transition
lines will not be reported, but only drawn in Figure \ref{fig:phdi.2+4} for 
completeness.

\begin{figure}
\includegraphics[scale=1.0]{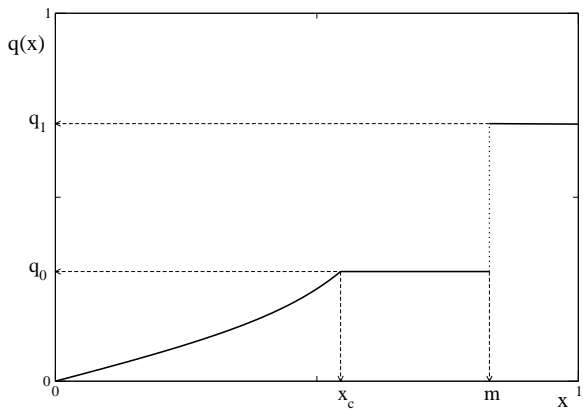}
\caption{Schematic form of the order parameter function $q(x)$ in the
1-FRSB phase.
}
\label{fig:qx-1frsb}
\end{figure}

{\it The One-Full Replica Symmetry Breaking phase} (1-FRSB).
The analysis of the instability of the 1RSB solution reveals that in order to 
stabilize the phase above the line where the replicon (\ref{eq:1rsb-stab})
vanishes a non-zero $q_0$ would be needed. However in the absence
of external fields the order parameter function must vanish as $x\to 0$, 
and hence a 1RSB solution is not possible. On the other the different location
of the static and dynamic instability lines suggests that some sort of 
1RSB-like form must survive in the solution. The way out is to look for a
solution which below $q_0$ has a structure which vanishes as $x\to 0$. The 
most general form is an order parameter that has a discontinuity 
at $x=m$, is continuous below it and vanish for $x=0$:
\begin{equation}
\label{eq:1frsb-qx}
q(x) = \left\{ \begin{array}{ll}
  q_1   & \mbox{for $x>m$} \\
  q_{0}(x)  & \mbox{for $x<m$} 
  \end{array}
\right.
\end{equation}
with $q(0) = 0$ and 
$\lim_{x\to m^-}q(x) = q_0 \not= \lim_{x\to m^+}q(x) = q(1) = q_1$, 
see Figure \ref{fig:qx-1frsb}.
That this is the correct {\it ansatz} also follows from the numerical solution
of the Parisi equations derived from stationarity of $f$ with respect
to $q(x)$ \cite{SomDup84,CriRiz02}.

The 1-FRSB equations are obtained by inserting the form 
(\ref{eq:1frsb-qx}) into the replica free energy 
(\ref{eq:rpl-f}) and imposing stationarity with respect to $q_0(x)$, 
$q_1$ and $m$. The resulting equations can be solved in term of
$m$-lines similarly to what done for the 1RSB case.
For the $p=4$  case the solution for the ``discontinuous'' part of $q(x)$ reads
\begin{eqnarray}
\label{eq:1-frsb-d}
\mu_4 &=& \frac{[1-y+my(1-t)]^4}{m^2 y (1-y)^2 (1-t)^3 (1+2t)} \\
\mu_2 &=& \frac{[1-y+my(1-t)]^2}{m^2 y (1-t)^3 (1+2t)} 
                   [y(1+t+t^2) - 3t^2] \nonumber
\end{eqnarray}
where 
\begin{equation}
\label{eq:t}
t = \frac{q_0}{q_1} = \frac{1+y - 2z(y)}{4z(y) - 3 - y}
\end{equation}
and $q_1 = (1-y) / [1-y + m y(1-t)]$. The ``continuous'' part of $q(x)$ is
given by
\begin{equation}
\label{eq:1-frsb-c}
q = \int_{0}^{q} dq' [\mu_2 + 3 \mu_4 q'^2] \, \chi(q')^2, 
\quad 0\leq q\leq q_0 = tq_1
\end{equation}
where
\begin{equation}
\label{eq:chi}
\chi(q) = 1 - q_1 + m (q_1 - q_0) + \int_{q}^{q_0} dq' x(q')
\end{equation}
For any fixed value of $m\in[0,1]$ these equations can be solved varying 
$y$ from $y_{\min} : x(y_{\min}) = 0$ (transition line to 1RSB phase) up to 
$y=1$ ($t(y=1) = 1$) where the difference between $q_1$ and $q_1$ vanishes. 
These lines are the continuation into the 1-FRSB phase
of the 1RSB $m$-lines. In particular the $m=1$ line represents the transition
between the 1-FRSB to the FRSB phase. For many aspects this transition is 
similar to the transition between the PM and the 1RSB phases, 
indeed $q_1$ jumps discontinuously from a null value (FRSB) to a finite value 
(1-FRSB) however the discontinuity appears at $m=1$ so that the
thermodynamic quantities are continuous across the transition. 
The critical $m=1$-line ends at the
point where $q_1=q_0$, which for $p=4$ is
\begin{equation}
q_1 = q_0 = \frac{1}{4},\ 
\mu_4 = \left(\frac{4}{3}\right)^{4},\
\mu_2 = \frac{32}{27}
\end{equation}
From this end-point on the transition between the FRSB 
and the 1-FRSB phases can only take place without a jump in in order parameter
function. The value of $t$, the ratio between $q_1$ and $q_0$, increases
along the $m$-lines as one moves away from the transition line with the 1RSB 
phase, and the lines terminates when $t=1$ ($q_0=q_1$). The set of
all end-points for $m\in[0,1]$ defines the continuous critical line between 
the 1-FRSB and the FRSB phases, which for $p=4$ reads
\begin{equation}
\mu_4 = \frac{1}{m^2} \left(\frac{1+3 m}{3}\right)^{4},\
\mu_2 = \frac{2}{3} \left(\frac{1+3 m}{3m}\right)^{2},\
\end{equation}
On this line $q_1=q_0 = 1 / (1+3m)$ 
and $x_c \equiv x(q_0) = m$. It is worth to note that in going from the 
1-FRSB to the FRSB phase the solution changes from stable to marginally 
stable, and remains marginally stable in the whole FRSB phase.

The presence of a discontinuity in in the order parameter function leads to
a finite complexity so that static and dynamics calculations leads to 
different solutions being the first associated with states of  smallest 
(zero) complexity while the latter with states of largest complexity.
As a consequence the $m$-lines in the two cases are different, as shown in
Figure \ref{fig:phdi.2+4}. We shall not report the expression for the 
dynamical $m$-line, this will be done elsewhere. The inset of the figure shows 
the different transition lines between the FRSB and the 1-FRSB phases. 
The discontinuity, and hence the complexity, vanishes on the continuous 
transition line and two solutions coincide on this line and in the whole 
FRSB phase.

{\it The Full Replica Symmetry Breaking phase} (FRSB).
In this phase the order parameter function is continuous and given
by eq. (\ref{eq:1-frsb-c}) with $q_1=q_0$. By expressing (\ref{eq:1-frsb-c})
in term of $q(x)$, instead of $x(q)$, ad taking successive
derivatives whit respect to $x$ a power expansion of $q(x)$ about $x=0$
can be computed \cite{SomDup84,Sommers85}. 
For the $2+4$ model it turns out that $q(x)$ contains
only odd powers of $x$, the first of which are
\begin{equation}
q(x) = \frac{\mu_2^{3/2}}{3\mu_4}\, x + 
       \frac{\mu_2^{7/2}}{6\mu_4^4}\,x^3 +
       \frac{13 \mu_2^{11/2}}{72\mu_4^3}\,x^7 + \cdots
\end{equation}
Using this expression one can show that as the PM-FRSB transition
line is approached from above $\tau = \mu_2 - 1 \to 0^+$ then
both $q_0 = q(x_c)$ and $x_c$ vanishes linearly with $\tau$ as
\begin{equation}
q_0 = \frac{\tau}{2} + O(\tau^2), \quad
x_c = \frac{3\mu_4}{2}\, \tau + O(\tau^2), \quad \tau\to 0^+.
\end{equation}
so that the transition between the PM and the FRSB phases takes place 
continuously without any jump in the order parameter function.

{\it Conclusions.} To summarize in this Letter we have provided 
the full phase diagram of the spherical $2+p$ spin glass model with $p\ge 4$.
Despite its simplicity the model has a rather rich phase. 
Indeed not only it presents a 1RSB phase similar to that
of the spherical $p$-spin spin glass model and a FRSB phase similar to that 
of the SK model, but it also posses a completely
new phase with an order parameter made of a continuous part for
$x<m\leq 1$ much alike the FRSB order parameter and a discontinuous jump
at $x=m$ resembling the 1RSB case. To emphasize its mixed nature
this phase, which separates the FRSB phase and the 1RSB phase, has been 
called 1-FRSB.
For may aspects the 1-FRSB phase is similar to the 1RSB, in particular 
it is stable, at difference with discrete models as
the SK model where a solution of this form is always unstable
\cite{CriLeuParRiz04},  and has a finite complexity 
counting metastable states that are strict minima
of the free energy landscape. 

The transition between the FRSB and the 1-FRSB phase can be either 
continuous (for $\mu_2$ large enough) or discontinuous. In the first case
the transition line is the continuation of the discontinuous transition
between the PM and the 1RSB phases and as for the latter the discontinuity 
appears at $m=1$. The presence of finite complexity in the 1-FRSB phase
makes the static and dynamics transition different.
The two lines join together at the end-point where the discontinuity 
at $m=1$ in the order parameter function vanishes. From this point on the 
transition between the FRSB phase and the 1-FRSB can only take place 
continuously with the discontinuity of the 1-FRSB phase which vanishes 
at the transition.
Along the the continuous transition line the complexity vanishes and the
static and dynamic approaches lead to the same results, in agreement 
with the conjecture made in Ref. \onlinecite{CriLeuParRiz04a}
that the FRSB phase has vanishing complexity.

In conclusion we believe that this is a rather promising model since not only 
it can be fully solved, but it possess different phases which can be
fully analyzed. Moreover it posses an interesting transition between
two different glassy phases, similar to what found in some 
colloidal suspensions \cite{DawetAl01}.


\begin{thebibliography}{99}
\bibitem{note1}
  The most general form of a the order parameter function in the 
  1RSB phase is $q(x) = q_0 \theta(m-x) + q_1\theta(x-m)$.
  However in the absence of external fields $q_0=0$.
\end{thebibliography}
\end{document}